\documentclass[pre,twocolumn,groupedaddress,showpacs]{revtex4}
\usepackage{graphicx,epsfig,amsmath}

\begin{document}

\title{Crystallization of medium length 1-alcohols\\ in mesoporous silicon: An X-ray diffraction study}

\author{Anke Henschel}
\author{Patrick Huber}
\email{p.huber@physik.uni-saarland.de}
\author{Klaus Knorr}
\email{knorr@mx.uni-saarland.de} \affiliation{Faculty of Physics and Mechatronics Engineering,
Saarland University, D-66041 Saarbr\"ucken, Germany\\}

\date{\today}

\begin{abstract}
The linear 1-alcohols n-C$_{\rm 16}$H$_{\rm 33}$OH, n-C$_{\rm 17}$H$_{\rm 35}$OH, n-C$_{\rm 19}$H$_{\rm 37}$OH have been imbibed and solidified in lined up, tubular mesopores of silicon with 10~nm and 15~nm mean diameters, respectively. X-ray diffraction measurements reveal a set of six discrete orientation states (''domains'') characterized by a perpendicular alignment of the molecules with respect to the long axis of the pores and by a four-fold symmetry about this direction, which coincides with the crystalline symmetry of the Si host. A Bragg peak series characteristic of the formation of bilayers indicates a lamellar structure of the spatially confined alcohol crystals in 15~nm pores. By contrast, no layering reflections could be detected for 10~nm pores. The growth mechanism responsible for the peculiar orientation states is attributed to a nano-scale version of the Bridgman technique of single-crystal growth, where the dominant growth direction is aligned parallelly to the long pore axes. Our observations are analogous to the growth phenomenology encountered for medium length n-alkanes confined in mesoporous silicon (Phys.~Rev.~E~\textbf{75},~021607~(2007)) and may further elucidate why porous silicon matrices act as an effective nucleation-inducing material for protein solution crystallization.
\end{abstract}


\maketitle



Molecular ensembles condensed into mesopores usually show melting temperatures reduced with respect to the bulk state \cite{Christenson, AlbaSim}. The close packed high-symmetry crystal structures of small globular molecules such as the fcc structure of the heavy rare gases or the hcp structure of N$_{\rm 2}$ and CO are conserved in pore confinement \cite{Huber99}. For the medium length n-alkanes the situation is different. These molecules form layered crystals \cite{Mueller1932}. Within the layers the molecules are tightly packed side-by-side, thereby forming a quasi-hexagonal 2D array.  In the so-called "crystalline, C" low-temperature phase the molecules are ordered with respect to rotations about the molecule axis. This leads to a long-range azimuthal order of the herringbone-type, accompanied by an uniaxial distortion of the lattice with respect to the hexagonal reference. Furthermore  the all-trans zig-zag chains of the C-atoms of lateral neighbours lock-in resulting in well defined collective tilts of the molecule axis with respect to the normal of the layers, the tilt angle is zero for alkanes with an odd number of C-atoms and of the order of 30~deg for even-numbered alkanes. Close to the melting temperatures mesophases (rotator R phases) appear with partial or complete rotational disorder, reduced uniaxial distortion and - in case of even alkanes - reduced tilt angles \cite{Sirota1, Sirota2, Doucet1, Doucet2, Dirand, Ewen1980}.

When confined in porous varieties of silica (e.g. Vycor) with pore diameters of a few nm, melting and the C-R transition temperature are reduced, the layering reflections are absent, the temperature range over which the R-phases are stable is increased, and the odd-even difference is lifted \cite{Huber1, Huber2}. Altogether the effects of pore confinement appear plausible. The pore networks of the silica substrates are highly random, the pore solid is subject to random strains which in turn couple to the rotational and translational degrees-of-freedom \cite{Hoechli1990}. Thereby the disordered phases are favored and the odd-even effect is blurred.

\begin{figure}[hbt]
\includegraphics[scale=0.36]{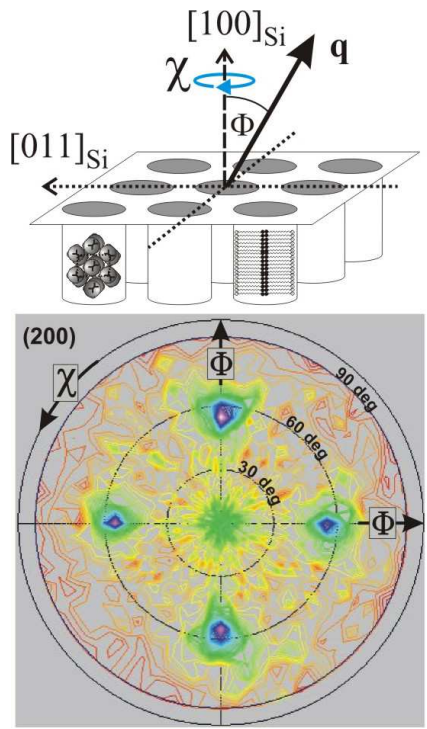}
\caption{\label{C19Polefigure} \small{(Color online) X-ray diffraction pole figure of the (20) Bragg reflection of C19OH confined in mesoporous silicon at 298~K. As illustrated in the upper panel, the orientation of the sample sheet with respect to the scattering vector \textbf{q} is specified by the polar angle $\Phi$ and the azimuth $\chi$. The two orientational domains of the molecules about the $\chi$ axis are schematically sketched in the figure: The orientation of the long axes of the molecules about the $\chi$-axis differ by 90~deg resulting in a view on the lateral herringbone-type ordered in-plane structure of the alcohols for the first domain (left pore) and a side view on the lamellar order of the chains for the second domain (right pore). Note, the long axes of the molecules in both $\chi$-domains are perpendicular to the long pore axis, which coincides with the [100] direction of the Si host.}}
\end{figure}

Recently we have reported x-ray diffraction results on the n-alkanes Cn, $n=16,17,19,25$ ($n$ stands for the number of C-atoms) in a porous Si (100) sheet with a pore diameter of 10~nm \cite{Henschel}. In this mesoporous substrate the pores are lined up, perpendicular to the sheet. In most respects the results were found to be analogous to those obtained in mesoporous silica, but the layering reflections do now show up and the crystal lattice of the alkanes has a well defined set of orientation states ("domains") with respect to the Si lattice. In particular, the long axes of the molecules are arranged perpendicular to the long axes of the pores, similarly as has been reported for the crystallization of folded polymer chains forming lamellae in aligned tubular alumina pores \cite{Steinhart}.

The present article deals with the corresponding \mbox{1-alcohols} C$n$OH, $n=16,17,19$ in porous Si \cite{Canham1995}. In the bulk state, the structural and thermodynamic behavior of C$n$ and C$n$OH are closely related as far as the phase sequence and the odd-even effect is concerned, the main difference being the fact that the monolayers of the alkanes are replaced by tail-to-tail bilayers in which the sublayers are coupled by H-bonds forming an O$-$H$\cdot\cdot$O chain \cite{Sirota, Ventola, Abrahamsson}.

The preparation of the substrate and the x-ray diffraction experiment has been described in Ref. \cite{Henschel, Huber2007}. The in-plane diffraction patterns of all three alcohols investigated are basically identical. They are dominated by the fundamental reflections of the quasi-hexagonal in-plane lattice which are indexed (11), (1-1), (20) in terms of the rectangular 2D lattice with the unit mesh containing two molecules. The lattice parameters, $a$ and $b$, can be extracted from the peak positions and converted into the uniaxial distortion $D$ with respect to hexagonal reference lattice, $D=1-a/b\sqrt3$, and the area per molecule $A=ab/2$. The scattering vector \textbf{q} forms a polar angle $\Phi$ with the pore axis (=sheet normal) and an azimuth $\chi$ about the pore axis - see also Fig.~1. The origin of the $\chi$ is chosen such that for $\chi$=0, \textbf{q} is along the [011] direction of the Si lattice. The fundamental triple can only be observed for $\Phi$ being close to a multiple of 60~deg and $\chi$ being close to a multiple of 90~deg. As discussed in detail in Ref.~\cite{Henschel}, this observation calls for six domain states, and we have argued that the discrete $\Phi$-values result from a Bridgman type selection process of the propagation of the solidification front along the pores, whereas the texture with respect to $\chi$ signals epitaxy on pore walls formed by (011) facets of the host lattice.

\begin{figure}[bt]
\includegraphics[scale=0.35, angle=90]{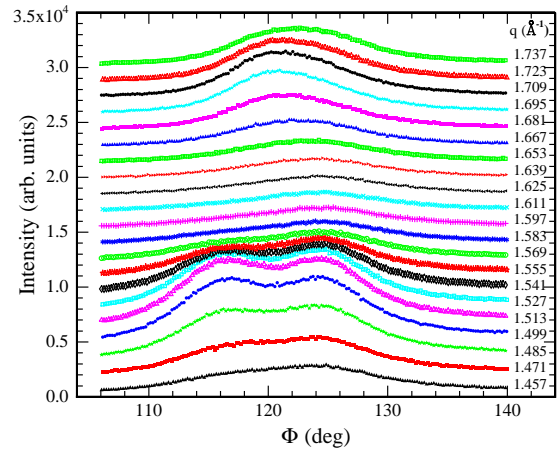}
\caption{\label{C17rockingscan} \small{(Color online) A series of $\Phi$-"rocking" scans on C17OH crystallized in mesoporous silicon at 245~K. The scattering vector \textbf{q} lies in the $\chi$=90~deg-plane. $\left|\textbf{q}\right|$ is held constant along an individual scan at a value specified in the figure. The three peaks observed are the fundamental triple mentioned in the text.}}
\end{figure}

\begin{figure}[bt]
\includegraphics[scale=0.48]{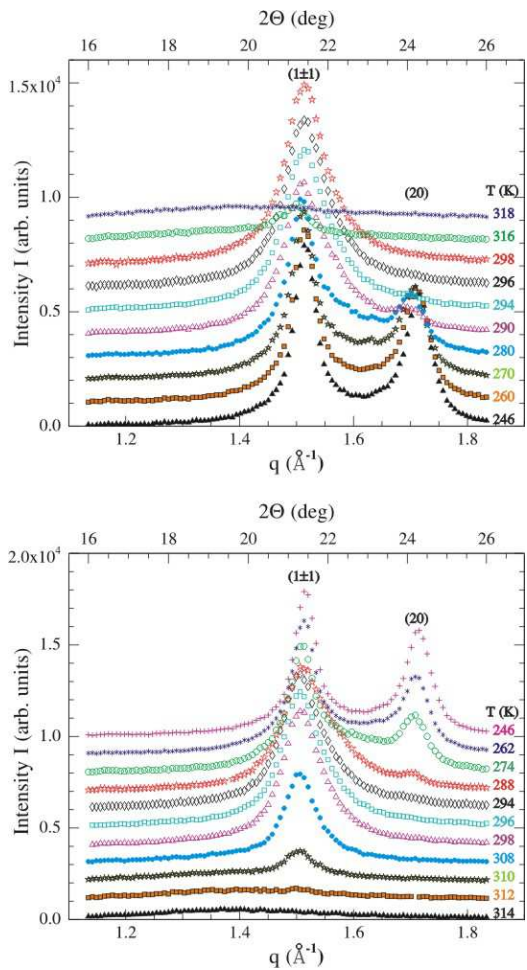}
\caption{\label{C17radialscans} \small{(Color online) A series of radial scans on C17OH solidified in mesoporous silicon showing the (11)/(1-1) and the (20) reflections at selected temperatures $T$ indicated in the figure, both on cooling and heating through the melting/freezing and the C-R transition. The scattering vector is parallel to the pore axis ($\Phi$=0). }}
\end{figure}

The formation of the domain states is apparent from Figs.~1 and~2. Fig.~1 is a pole figure of the (20)~reflection. The fourfold symmetry with respect to $\chi$-rotations about the pore axis is evident. In Fig.~2 we show the fundamental triple by means of $\Phi$ rocking scans in the $\chi$=90~deg-plane for a series of different Bragg angles 2~$\Theta$ and hence moduli of the scattering vector \textbf{q}. As can be seen the triple is centered at $\Phi$=120~deg. The three peaks represent the three $\Phi$-domains ($\Phi$=0~deg,$\pm$60~deg) which obviously have about equal statistical weights.  The changes of the diffraction pattern with temperature, both on cooling and heating, are illustrated by the scans of Fig.~3 where \textbf{q} is parallel to the pore axis ($\Phi$=0). Such a radial scan hits the (20) peak directly, the (11) and (1-1) peaks are slightly off the scan path, but their intensity is picked up because of the finite mosaic width. The freezing and melting temperatures, $T_{\rm f}$ and $T_{\rm m}$, can be determined from the onset of Bragg intensities on cooling and their disappearance on heating, respectively. The peak splitting signals the transition from the R to the C phase which also shows some thermal hysteresis between cooling and heating. The transition temperatures are collected in Table I. The low-$T$ phase shows (21)-reflections characteristic of a herringbone type orientational order and has in-plane lattice parameters $a$ and $b$ that are within experimental error identical to those of prototypic orthorhombic C phase of the alkanes which in turn is closely related to the fully ordered, low tilt monoclinic $\beta$-phase of the odd numbered 1-alcohols \cite{Ventola, Yamamoto}. This justifies calling the low-T phase "C".  Close to $T_{\rm f}$/$T_{\rm m}$ the fundamental triple merges into a single peak which points to the hexagonal in-plane lattice, known from the rotator phase R$_{\rm II}$. On the other hand the peaks of the mesophases are always asymmetric which could mean that there is some non-resolved peak splitting due to residual distortions with a magnitude similar to what has been observed in the rotator R$_{\rm V}$ and R$_{\rm IV}$ phases of the bulk alkanes and the analogous phases of the alcohols. Altogether the in-plane metric observed does not tolerate tilt angles of the order of 30~deg that have been observed in C phases of the bulk even-numbered alkanes and alcohols. In pores the tilt angle practically vanishes, not only for the odd-numbered molecules, but also in C16OH. The peculiar texture observed not only means that the layer normal, but also that the molecules are oriented perpendicular to the pore axis.

\begin{table}
\begin{tabular}{c|c|c|c|c|c|c}
alcohol & $T_{\rm f/m}^{\rm cooling}$ & $T_{\rm f/m}^{\rm heating}$ & $T_{\rm R/C}^{\rm cooling}$ & $T_{\rm R/C}^{\rm heating}$ & $T_{\rm f/m}^{\rm Bulk}$ & $T_{\rm R/C}^{\rm Bulk}$ \\
\hline \hline
C16OH & 304\,K & 312\,K & 290\,K & 293\,K & 322\,K & 316\,K\\
C17OH & 310\,K & 318\,K & 294\,K & 298\,K & 326\,K & 314\,K\\
C19OH & 320\,K & 327\,K & 306\,K & 313\,K & 334\,K & 324\,K\\
\end{tabular}
\caption{\label{Tab:FreezingT} \small{Table of melting and freezing temperatures $T_{\rm f/m}$ and phase transition temperatures $T_{\rm R/C}$ on heating and cooling for n=16, 17, 19. The bulk data are given for comparison \cite{Sirota}.}}
\end{table}

Layering reflections have been searched for, in particular in scans with \textbf{q} perpendicular to the pores, but except for a very weak (001) reflection, no higher layering reflections have been observed, quite in contrast to the corresponding alkanes. Based on space filling arguments we do in fact conclude from the mass uptake of the pores that the alcohols are arranged in tightly packed parallel layers, but that there is a sizeable mean square displacement of the molecules in the $z$-direction perpendicular to layers (larger than the 2~\r A thick interlayer gap), such that the electron density contrast between the layers and the interlayer gaps, on which the intensity of such reflections relies, is washed out. The absence of the layering reflections in the alcohols may be related to the fact that the thickness of the bilayers (e.g. 4.8~nm for C17OH) is already getting close to the pore radius, on the other hand one might have thought that the strong H-bonds within the bilayer could help to reduce the $z$-excursions of molecules relative to the situation encountered for the alkanes. Obviously this is not the case.

\begin{figure}[hbt]
\includegraphics[scale=0.4]{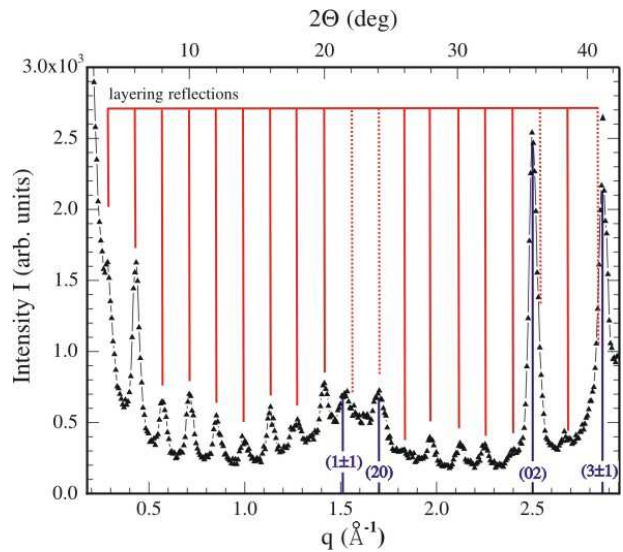}
\caption{\label{C17pSi15} \small{(Color online) Radial scan on C16OH solidified in mesoporous silicon. The scattering vector is perpendicular to the pore axis ($\Phi$=90~deg). The reflection comb characteristic of the equidistant Bragg peaks due to bilayer formation is included as a guide for the eye. Solid lines of the comb mark clearly observed bilayer Bragg peaks, whereas dotted lines indicate allowed bilayer Bragg peak positions, which, within our $q$-resolution, coincide with the in-plane reflections (11)/(1-1), (20), (31), and (3-1), respectively.}}
\end{figure}
The diffraction experiments have been repeated after the pore walls had been oxidized by a treatment with H$_{\rm 2}$O$_{\rm 2}$. See Ref. \cite{Henschel}. The melted alcohols are still sucked into the pores, but are pushed out of the pores upon solidification. The solid then forms epitaxial layers on the surface of the substrate with bulk properties, analogous to what has been observed for the alkanes.

In an additional preparation step, the oxide on the pore walls has been removed by etching with HF. This procedure increases the pore diameter to about 15~nm and also reduces the roughness of the pore walls \cite{Kumar2008}. Now the pore solid is again stable. The in-plane diffraction patterns are practically identical to those on the original samples, apart from the fact that the Bragg peaks are somewhat sharper and that the transition temperatures are slightly higher. However, in radial scans with \textbf{q} perpendicular to the pore axis the layering peaks show up - see Fig. \ref{C17pSi15}. Note, due to the parallel alignment of the layers to the long pore axes and the 90~deg texture about the $\chi$ axis, three out of six domains and hence half of the entire crystallized sample contributes to the intensity of the layering peaks in this scan geometry rendering them observable up to the 18$^{\rm th}$ order. The bilayer spacings conform to the relation $d=(4 + 2.56\, n)$~\r A which has been established in case of extended, all-trans molecules and vanishing tilt. The coherence length derived from the width of the layering peaks is 14~nm, which is close to the pore diameter.

Finally, we would like to speculate that the recently reported advantages of mesoporous silicon matrices in order to induce and speed up bulk crystallization in protein solutions \cite{Chayen} may not only be attributable to an increased heterogeneous crystal nucleation in the pores and at the substrate surfaces \cite{Page2006}, but also, to some extent, to a Bridgman type alignment mechanism of the fast growing crystal nuclei. The anisotropic pore confinement may align the dominant growth direction of protein seeds within the pores parallel to the long axes of the pores, similarly as demonstrated here for n-alcohols. Once these fast growing nuclei reach the pore mouths, they can induce fast, directed bulk protein crystallization in the bulk solution surrounding mesoporous Si.

\begin{acknowledgments}
We thank P. Leibenguth for taking the pole figure. The work has been supported by the German Research Foundation (DFG) via the Collaborative Research Centre (SFB) 277, Saarbr\"ucken.
\end{acknowledgments}


\begin{thebibliography}{99}
\bibitem{Christenson} H.K. Christenson, J. Phys. Condens. Mat.
\textbf{13}, R95 (2001); K. Knorr, P. Huber, and D. Wallacher, Z. Phys. Chem. \textbf{222}, 257 (2008).
\bibitem{AlbaSim} C. Alba-Simionesco, B. Coasne, G. Dosseh, G. Dudziak, K.E. Gubbins, R. Radhakrishnan, and M.G. Sliwinska-Bartkowiak, J. Phys. Condens. Mat.
\textbf{18}, R15 (2006).
\bibitem{Huber99} P. Huber and K. Knorr, Phys. Rev. B \textbf{60}, 12657 (1999); P. Huber and K. Knorr, Mater. Res. Soc. Symp. Proc. 876E, R3.1 (2005) and cond-mat/0508683.
\bibitem{Mueller1932} A.~M\"uller, Proc. Roy. Soc. A \textbf{138}, 514 (1932).
\bibitem{Sirota1} E. B. Sirota, H. E. King, Jr. , D. M. Singer, and Henry H. Shao, J. Chem. Phys. \textbf{98}
(1993).
\bibitem{Sirota2} E. B. Sirota and A. B. Herhold, Science \textbf{283}, 529
(1999).
\bibitem{Doucet1} J. Doucet, I. Denicolo, and A. Craievich, J. Chem. Phys. \textbf{75}, 1523
(1981).
\bibitem{Doucet2} J. Doucet, I. Denicolo, A. Craievich, and A. Collet, J. Chem. Phys. \textbf{75}, 5125 (1981).
\bibitem{Dirand} M. Dirand, M. Bouroukba, V. Chevallier, D. Petitjean, E. Behar, and V. Ruffier-Meray, J. Chem. Data \textbf{47} 115
(2002).
\bibitem{Ewen1980} B. Ewen, G.~R. Strobl, and D.~Richter, Faraday Disc. \textbf{69}, 19 (1980).
\bibitem{Huber1} P.~Huber, D.~Wallacher, J.~Albers, and K.~Knorr, Europhys. Lett. \textbf{65}, 351 (2004).
\bibitem{Huber2} P.~Huber, V.P.~Soprunyuk, and K.~Knorr, Phys.~Rev.~E~\textbf{74},~031610~(2006).
\bibitem{Hoechli1990} U. H\"ochli, K. Knorr and A. Loidl, Adv. Phys. \textbf{39} (1990) 405.
\bibitem{Henschel} A.~Henschel, T.~Hofmann, P.~Huber, and K.~Knorr, Phys. Rev. E \textbf{75},021607 (2007).
\bibitem{Steinhart} M. Steinhart, P. G\"oring, H. Dernaika, M. Prabhukaran, U. G\"osele, E. Hempel, and T. Thurn-Albrecht, Phys. Rev. Lett. \textbf{97}, 027801 (2006); E. Woo, J. Huh, Y.G. Jeong, and K. Shin, Phys. Rev. Lett. \textbf{98}, 136103 (2007).
\bibitem{Canham1995} L.T.~Canham, Adv. Mat. \textbf{7}, 1033 (1995).
\bibitem{Sirota} E.B. Sirota and X.Z. Wu, J. Chem. Phys. \textbf{105}, 7763 (1996).
\bibitem{Ventola} L. Ventola, M. Ramirez, T. Calvet, X. Solans, M.A. Cuevas-Diarte, P. Negrier, D. Mondieig, J.C. van Miltenburg, and H.A.J. Oonk, Chem. Mater. \textbf{14}, 508 (2002).
\bibitem{Abrahamsson} S. Abrahamsson, G. Larsson, and E. von Sydow, Acta Cryst. \textbf{13}, 770 (1960).
\bibitem{Huber2007} P. Huber, S. Gruener, C. Schaefer, K. Knorr, and A.V. Kityk, Europ. Phys. J. Spec. Top. \textbf{141}, 101 (2007).
\bibitem{Yamamoto} T. Yamamoto, K. Nozaki, and T. Hara, J. Chem. Phys. \textbf{92}, 631 (1990).
\bibitem{Kumar2008} P. Kumar, T. Hofmann, K. Knorr, P. Huber, P. Scheib, and P. Lemmens, J. Appl. Phys. \textbf{103}, 024303 (2008).
\bibitem{Chayen} N.E. Chayen, E. Saridakis, R. El-Bahar, Y. Nemirovsky, J. Mol. Biol. \textbf{312}, 591 (2001); N.E. Chayen, E. Saridakis, and R.P. Sear, Proc. Nat. Acad. Sc. \textbf{103}, 597 (2006); D. Frenkel, Nature \textbf{443}, 641 (2006).
\bibitem{Page2006} A.J. Page and R.P. Sear, Phys. Rev. Lett. \textbf{97}, 065701 (2006); S. Stolyarova, E. Saridakis, N.E. Chayen, and Y. Nemirovsky, Biophys. J. \textbf{91}, 3857 (2006).
\end{thebibliography}
\end{document}